\documentclass[twocolumn,pra,showpacs,superscriptaddress]{revtex4}
\usepackage{amssymb}
\usepackage{amsmath}
\usepackage{graphicx}
\usepackage{subfigure}
\usepackage{natbib}
\usepackage{epsfig}
\usepackage{amsfonts}
\usepackage{mathrsfs}
\usepackage{CJK}
\usepackage[toc,page,title,titletoc,header]{appendix}

\begin{document}

\title{Single-photon scattering on a strongly dressed atom}
\author{Z. H. Wang}
\affiliation{Institute of Physics, Chinese Academy of Sciences, Beijing, 100190, China}
\author{Yong Li }
\affiliation{Beijing Computational Science Research Center, Beijing, 100084, China}
\author{D. L. Zhou }
\affiliation{Institute of Physics, Chinese Academy of Sciences, Beijing, 100190, China}
\author{Chang-pu Sun }
\affiliation{Institute of Theoretical Physics, Chinese Academy of Sciences, Beijing,
100190, China}
\author{Peng Zhang }
\email{pengzhang@ruc.edu.cn}
\affiliation{Department of Physics, Renmin University of China, Beijing, 100872, China}

\begin{abstract}
We develop the generalized rotating-wave approximation (GRWA) approach
(Phys. Rev. Lett. \textbf{99}, 173601 (2007)) to study the single-photon
scattering on a two-level system (TLS) with arbitrarily\textbf{\ }strong
coupling to a local mode in a one-dimensional (1D) coupled resonator array.
We calculate the scattering amplitudes analytically, and the independent
numerical results show that our approach works well in a broad parameter
region. Especially, when the resonator mode is far-off resonance with the
TLS, we obtain more reasonable results than the ones from the standard
adiabatic approximation. Our approach is further extended to the cases with
a $1$D resonator array strongly coupled to more than one TLSs.
\end{abstract}

\maketitle


\section{introduction}

Recently, many attentions have been paid to the photon transport in a
low-dimensional array of coupled resonators~\cite%
{fan07,zhou08a,zhou08b,zhou08bb,gong08,song08,zhou08n,Jaouen09,liao09,shi09b,hu09,longo09,guo09,liao10,jang10,xu10,alexanian10,schmitteckert10,lu10, zhao11,Hollenberg11,Busch11,chang11,zhou09a,roy11}
or a one-dimensional (1D) wave guide~\cite%
{ji11,roy11b,zhou09a,roy11,fan05,fan07b,fan07c,fan09,fan09b,zhou08,guillermo09,
guillermo092,fan11n,lujing09,law09,dong09,shi09, shi10,jang10b,witthaut10,liao10c,baranger10,fan10,roy10,hoi10,fan10b}%
, which plays crucial roles in the realization of all-optical quantum
devices. In these systems, the two-level or multi-level devices coupled to
the resonators or wave guides can be used as quantum switches to control the
scattering or transport of the photons.

Up to now, the single-photon scattering amplitudes in the 1D resonator
arrays or wave guides coupled to a single two-level system (TLS)~\cite%
{fan07,zhou08a,zhou09a,dong09,fan05,fan09,fan09b,xu10,jang10,jang10b,fan10,lu10,zhao11}
or a single three-level system~\cite{gong08,witthaut10,law09,hoi10} or
multi quantum devices~\cite{zhou08b,zhou08bb,chang11,hu09,zhou08,guillermo09,guillermo092,fan11n,lujing09,guo09}
have been well-investigated by different authors.The relevant multi-photon
scattering amplitudes~\cite%
{shi09,shi09b,shi10,zhou09a,roy11,baranger10,roy10,schmitteckert10,fan10b,Busch11,ji11,roy11b}
are also studied. To our knowledge, all these studies are based on the
rotating-wave approximation which is applicable under the conditions that
the frequencies of the two-level or multi-level system are very close to the
photon frequencies in the resonators or wave guides, and the intensities of
the coupling between the two-level or multi-level system and the photons are
much smaller than their frequencies, namely, in the weak-coupling limit.

In the experiments, the $1$D resonator arrays or wave guides can be realized
with photonic crystals~\cite{faraon07,noda03}, superconducting transmission
line resonators~\cite{fan05,zhou08bb,liao10,liao09,Schoelkopf04,hoi10} or
other solid devices, while the two-level or multi-level systems can be
implemented with either natural atoms or the solid-state artificial atoms.
In the hybrid systems of solid-state devices, it has been predicted that~%
\cite{strong15,strong16,strong18,strong19} one can realize ultra-strong
TLS-photon coupling with intensities comparable or even higher than the
photon frequencies. On the other hand, the frequency of the solid-state TLS
can also be controlled easily in a broad region. Therefore, in the quantum
network based on solid-state devices it is possible to reach the
strong-coupling and far-off resonance regime where the conditions of
rotating-wave approximation are violated. In these parameter regions, we
need to develop new methods for the scattering between the flying photon and
the TLS, and then investigate the possible new effects given by the strong
TLS-photon coupling to the photonic transport.

In this paper, we study the single photon scattering on the TLS, which
strongly couples to the local mode of the resonators. Beyond the
rotating-wave approximation, we develop an analytical approach for the
approximate calculation of the single-photon scattering amplitudes in a 1D
single-mode-resonator array coupled to a single TLS. Our method is based on
the generalized rotating-wave approximation (GRWA) ~\cite{irish} developed
by Irish for the system with a single TLS coupled to a single-mode bosonic
field.

With our GRWA approach we obtain the single-photon scattering amplitudes
under the condition that the photon hopping between different resonators is
weak enough. We will show that our approach works significantly well in a
very broad parameter region, including the region where the rotating-wave
approximation is applicable and the one where the TLS-photon coupling is
strong while the frequency of the TLS is close to or smaller than the photon
frequency.

Especially, in the far-off resonance region where the TLS frequency is much
smaller than the flying photon, the standard adiabatic approximation does
not work well in the current hybrid system, while our approach still provides
good results. On the other hand, when photon-TLS coupling is strong enough,
our GRWA approach shows that the single-photon scattering by the TLS becomes
equivalent to the transport of a single photon in a 1D resonator array in
which the frequency of a certain resonator is shifted. Then our results are
furthermore simplified and one can make reasonable qualitative estimations
for the photon transport characters, even without the quantitative
calculations. We also show that, our GRWA approach can be generalized to the
systems with a 1D single-mode-resonator array coupled with more than one
TLS.

This paper is organized as follows: In Sec.~\ref{sec2},
we develop the GRWA approach for the hybrid system of a 1D resonator array
coupled to a single TLS. In Sec.~\ref{sec3}, we analytically calculate the
single-photon scattering amplitudes in such a system with our GRWA approach
and compare our results with the numerical results. In Sec.~\ref{sec4}, we
discuss the single-photon scattering problem in the cases of strong
TLS-photon coupling. In Sec.~\ref{sec5}, we show the GRWA approach in the
system with more than one TLS. There are several discussions and conclusions
in Sec.~\ref{sec6}.

\section{The GRWA for the TLS-coupled $1$D resonator array}

\label{sec2}

In this paper we consider the transmission of a single photon in a $1$D
single-mode-resonator array coupled to a TLS which is located in a specific
resonator. To obtain the reasonable analytical results beyond the
rotating-wave approximation, in this section we will generalize the GRWA
approach proposed by Irish in Ref.~\cite{irish} for a single resonator
coupled to a TLS to our current hybrid system of resonator array. We will first show the
adiabatic approximation in our system for the cases where the frequency of
the TLS is much smaller than the photon frequency, and then develop the GRWA
approach as an improvement of the adiabatic approximation. For the reader's
convenience, in appendix A we introduce Irish's GRWA approach for the system
with a single resonator in the view of adiabatic approximation.

\subsection{The system and Hamiltonian}

\begin{figure}[tbp]
\includegraphics[bb=100 511 402 692,clip,width=6cm]{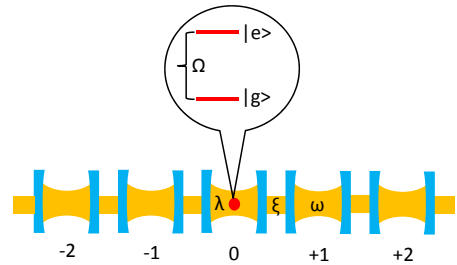}
\caption{(Color online) Schematic configuration for the hybrid system of a
1D resonator array interacting with a TLS. The frequency of the photon in
each resonator is $\protect\omega$, while the intensity of the photonic
hopping is $\protect\xi$. The TLS with frequency $\Omega$ is localized at
the $0$-th resonator and coupled to the photon with coupling strength $%
\protect\lambda$.}
\label{cavity}
\end{figure}

As shown in Fig.~\ref{cavity}, we consider a $1$D array of infinite
identical single-mode resonators with a TLS located inside a certain
resonator, which is marked as the $0$-th resonator in the array. We further
assume that the photons can hop between neighbor resonators. Then the total
system is modeled by the Hamiltonian $H=H_{C}+H_{A}+H_{I}$, where the
tight-binding Hamiltonian $H_{C}$ of the resonator array is
\begin{equation}
H_{C}=\omega \sum_{j=-\infty }^{+\infty }a_{j}^{\dagger }a_{j}-\xi
\sum_{j=-\infty }^{+\infty }(a_{j}^{\dagger }a_{j+1}+h.c.).  \label{hc}
\end{equation}%
Here $\omega $ is the frequency of the photons in the resonators, $\xi $ is
the hopping intensity or the inter-resonator coupling strength, $a_{j}$ and $%
a_{j}^{\dagger }$ \ are the annihilation and creation operators of the
photons in the $j$-th resonator respectively.

In this paper we assume the weak-hopping condition
\begin{equation}
\left\vert \xi \right\vert <<\omega  \label{wc}
\end{equation}%
is satisfied. Therefore the term $a_{j}^{\dagger }a_{j+1}^{\dagger }+h.c.$
has been neglected in our consideration. This condition also provides us a
small parameter $\xi /\omega $ which is very useful in the following
calculation.

The Hamiltonian $H_{A}$ of the TLS and the interaction $H_{I}$ between the
TLS and the photons in the $0$-th resonator are
\begin{equation}
H_{A}=\frac{\Omega }{2}\sigma _{z}  \label{ha}
\end{equation}%
and
\begin{equation}
H_{I}=\lambda \sigma _{x}(a_{0}^{\dagger }+a_{0})  \label{h11}
\end{equation}%
respectively. Here $\Omega $ is the energy spacing between the ground state $%
|g\rangle $ and the excited state $|e\rangle $ of the TLS, $\lambda $ is the
coupling intensity and the Pauli operators $\sigma _{z}$ and $\sigma _{x}$
are defined as $\sigma _{z}=\left\vert e\right\rangle \left\langle
e\right\vert -\left\vert g\right\rangle \left\langle g\right\vert $ and $%
\sigma _{x}=\left\vert e\right\rangle \left\langle g\right\vert +\left\vert
g\right\rangle \left\langle e\right\vert $.

\subsection{The adiabatic approximation for the TLS-coupled $1$D resonator
array}

In this and the next subsections, we develop the GRWA approach for the
TLS-coupled 1D resonator array. To this end, in this section we first
formally develop the adiabatic approximation for our system in the far-off
resonance region with $\Omega <<\omega $. We will show that, due to the
energy band structure of the photons, the adiabatic approximation \textit{%
cannot} be used in our hybrid system, even in the far-off resonance case.
However, as shown in the next subsection, we can also develop the GRWA
approach as an improvement of this formal adiabatic approach, and the
intrinsic problem of the adiabatic approximation in our system is naturally
overcome by our GRWA approach. Finally, our approach is applicable in a
broad parameter region, including the cases of $\Omega <<\omega $ and $%
\Omega \sim \omega $.

Under the far-off resonance condition $\Omega <<\omega $, the TLS is
considered to be the slowly-varying part of the system, while the 1D
resonator-array is the fast-varying one. Then we decompose the total
Hamiltonian $H$ as $H=H_{1}+H_{2}$ where $H_{1}=H_{C}+H_{I}$ is the
Hamiltonian of the fast-varying part together with the interaction and $%
H_{2}=H_{A}$ is the Hamiltonian of the slow-varying TLS.

The straightforward calculation shows that $H_{1}$ has the eigen-states
\begin{equation}
|\pm ,{\vec{n}}\rangle =U\prod_{k}\frac{1}{\sqrt{n(k)!}}A(k)^{\dagger
n(k)}|0\rangle \otimes |\pm \rangle .  \label{a1}
\end{equation}%
Here the TLS states $|\pm \rangle $ are the eigen-states of the operator $%
\sigma _{x}$ with eigen-values $\pm 1$, and $|0\rangle $ is the vacuum state
of the resonator array. The photon momentum $k$ can take any value in the
region $(-\pi ,\pi ],$ the creation operator $A(k)^{\dagger }$ for a photon
with momentum $k$ is given by
\begin{equation}
A(k)^{\dagger }=\frac{1}{\sqrt{2\pi }}\sum_{l=-\infty }^{+\infty
}e^{ikl}a_{l}^{\dagger },  \label{bigak}
\end{equation}%
and%
\begin{equation}
{\vec{n}=}(n(k_{1}),n(k_{2}),...)
\end{equation}%
is the set of all the numbers $n\left( k\right) $.

The operator $U$ in Eq.~(\ref{a1}) is determined by
\begin{eqnarray}
U^{-1}H_{1}U &=&\omega \sum_{j=-\infty }^{+\infty }a_{j}^{\dagger }a_{j}-\xi
\sum_{j=-\infty }^{+\infty }(a_{j}^{\dagger }a_{j+1}+h.c.)-\mathcal{C},
\notag  \label{ut2} \\
&&
\end{eqnarray}%
with $\mathcal{C}$ an unimportant c-number. This equation leads to the
result
\begin{equation}
U=\prod\limits_{j=-\infty }^{+\infty }\exp [\alpha _{j}\sigma
_{x}(a_{j}^{\dagger }-a_{j})]  \label{utt1}
\end{equation}%
with
\begin{equation}
\alpha _{j}=\frac{\lambda \omega _{1}}{2\xi ^{2}-\omega \omega _{1}}\left(
\frac{\xi }{\omega _{1}}\right) ^{|j|},  \label{alphaj1}
\end{equation}%
where
\begin{equation}
\omega _{1}=(\omega +\sqrt{\omega ^{2}-4\xi ^{2}})/2.  \label{omega11}
\end{equation}%
The results in Eqs.(\ref{alphaj1},\ref{omega11}) are derived in appendix B.
It is pointed out that the parameter $\alpha _{j}$ exponentially decays to
zero in the limit $|j|\rightarrow \infty $.

\begin{figure}[tbp]
\includegraphics[bb=50 173 536 552,clip,width=8.5cm]{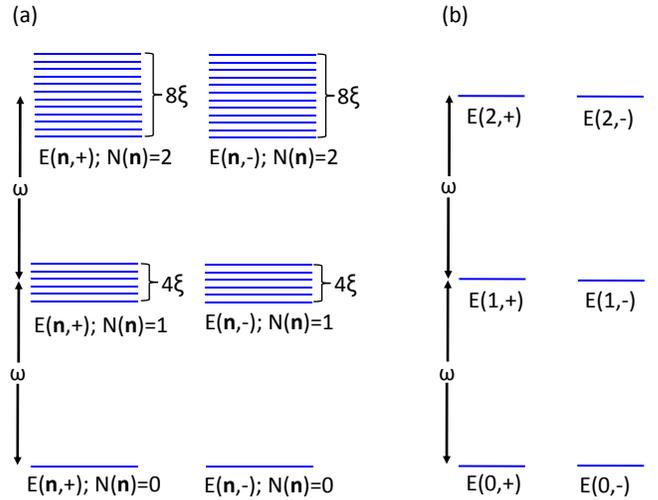}
\caption{(Color online) (a) The energy spectrum of the Hamiltonian $H_{1}$
of the 1D resonator-array and the interaction between the resonator-array
and the TLS. The low-excited spectrum has clear band structure. The
intra-band transitions are missed unreasonably in the adiabatic
approximation. (b) The energy spectrum of the Hamiltonian $H_{1}$ of a
single-mode bosonic field and the interaction between the single-mode
bosonic field and a TLS. In such a simple system there is no band structure
in the spectrum of $H_{1}$, and then the adiabatic approximation is
applicable when $\Omega<<\protect\omega$. }
\label{fig1}
\end{figure}

Obviously, the eigen-energy of $H_{1}$ with respect to the eigen-state $|\pm
,{\vec{n}}\rangle $ is
\begin{equation}
E(\pm ,{\vec{n}})=N({\vec{n}})\omega +2\xi \sum_{k}n(k)\cos k+\mathcal{C}
\label{a2}
\end{equation}%
with
\begin{equation}
N({\vec{n}})=\sum_{k}n(k).  \label{bign}
\end{equation}%
As shown in Fig.~\ref{fig1}(a), in the weak-tunneling case with $\xi
<<\omega $ the low-excited spectrum $E(\pm ,{\vec{n}})$ of the
eigen-energies of $H_{1}$ has a clear band structure. Each energy band
includes all the energy levels with the same total photon number $N(\vec{n})$
and different photon-momentum distribution $\vec{n}$. The $N$-th energy band
is centered at $N\omega $ with band width $4N\xi $. Therefore, in the low
excitation cases the inter-band energy gap has the same order with $\omega $.

The spirit of the adiabatic approximation~\cite{BOA1,BOA2,pengBOA} is that,
during the quantum evolution, the motion of the slowly-varying part of the
system would follow the motion of the fast-varying one. Then the quantum
state of the fast-varying 1D resonator array would be frozen in each
adiabatic branch with fixed quantum number $\vec{n}$. Mathematically
speaking, in the adiabatic approximation for our system, all the $H_{2}$%
-induced quantum transitions between the states $|\pm ,\vec{n}\rangle $ and $%
|\pm ,\vec{n^{\prime }}\rangle $ with $\vec{n}\neq \vec{n^{\prime }}$ are
neglected. This treatment leads to the approximate eigen-states of the
Hamiltonian $H$ as
\begin{equation}
|\Psi _{\pm ,\vec{n}}\rangle =\frac{1}{\sqrt{2}}\left( |+,\vec{n}\rangle \pm
|-,\vec{n}\rangle \right) .  \label{a6}
\end{equation}

Now we point out that, the adiabatic approximation is \textit{not} a
reasonable approximation for our present system, even in the case of $\Omega
<<\omega $. That is because, the energy spectrum of $H_{1}$ has a band
structure, each band includes all the states $|\pm ,\vec{n}\rangle $ with
the same total photon number $N(\vec{n})$. On the other hand, in the
adiabatic approximation all the $H_{2}$-induced transitions between the
eigen-states of $H_{1}$ with different quantum number ${\vec{n}}$ are
neglected. It means that, all the inter-band transitions and the intra-band
transitions between the states $|+,\vec{n}\rangle $ and $|-,\vec{n}^{\prime
}\rangle $ with ${\vec{n}}\neq \vec{n}^{\prime }$ are omitted. In the case
of $\Omega <<\omega $, the omission of inter-band transitions are reasonable
because the energy gap between different bands is of the order of $\omega $,
which is much larger than the intensity $\Omega $ of $H_{2}$. However, the
neglecting of the intra-band transitions are unreasonable because the energy
gaps between the levels in the same band can be arbitrary small.

As a comparison, we re-consider the simple system with a single-mode
bosonic field interacting with a TLS. In Fig.~2(b), we plot the spectrum of
the Hamiltonian
\begin{equation}
H_{\mathrm{Rabi}1}=\omega a^{\dagger }a+\lambda \sigma _{x}(a+a^{\dagger })
\end{equation}%
with $(a,a^{\dagger })$ the annihilation and creation operators of the
bosonic field. It is apparent that in such a simple case the spectrum of $H_{%
\mathrm{Rabi1}}$ does not have any band structure (shown in appendix A).
Therefore the adiabatic approximation~\cite{irish05} is applicable when $%
\Omega <<\omega $.

Although the adiabatic approximation cannot be used in our system, as shown
in the next subsection, we can still develop the GRWA approach for our
system with the help of the basis $\left\{ |\Psi _{\pm ,\vec{n}}\rangle
\right\} $. Furthermore, in our GRWA approach the above intrinsic problem of
the adiabatic approximation is naturally overcome.

\subsection{The GRWA approach for the TLS-coupled 1D resonator
array}

Now we develop the GRWA approach for the system of the TLS-coupled 1D
resonator array. Similar to the original GRWA for the single-mode bosonic
field coupled to a TLS (appendix A), our general GRWA approach can also be
considered as an improvement of the adiabatic approximation.

In our GRWA approach the total Hamiltonian $H$ is approximated as $H_{G}$
which is defined as
\begin{eqnarray}
H_{G} &=&\sum_{{\vec{n}},{\vec{n}^{\prime }}}\sum_{\alpha =+,-}H_{{\vec{n}%
^{\prime }},\alpha }^{{\vec{n}},\alpha }|\Psi _{\alpha ,\vec{n}}\rangle
\langle \Psi _{\alpha ,{\vec{n}}}|\delta _{N({\vec{n}}),N({\vec{n}^{\prime }}%
)}  \notag \\
&&+\sum_{{\vec{n}},{\vec{n}^{\prime }}}H_{{\vec{n}^{\prime }},-}^{{\vec{n}}%
,+}|\Psi _{+,{\vec{n}}}\rangle \langle \Psi _{-,{\vec{n}^{\prime }}}|\delta
_{N({\vec{n}}),N({\vec{n}^{\prime }})-1}+h.c.,  \notag \\
&&  \label{htg}
\end{eqnarray}%
where the matrix elements $H_{{\vec{n}^{\prime }},\beta }^{{\vec{n}},\alpha
} $ are defined as
\begin{equation}
H_{{\vec{n}^{\prime }},\beta }^{{\vec{n}},\alpha }=\langle \Psi _{\alpha ,{%
\vec{n}}}|H|\Psi _{\beta ,{\vec{n}^{\prime }}}\rangle  \label{hme}
\end{equation}%
while the symbol $\delta _{a,b}$ is defined as $\delta _{a,b}=1$ for $a=b$
and $\delta _{a,b}=0$ for $a\neq b$.

Apparently, in the approximate Hamiltonian $H_{G}$ we only take into account
the quantum transitions between the states $|\Psi _{\alpha ,{\vec{n}}%
}\rangle $ and $|\Psi _{\beta ,{\vec{n}}^{\prime }}\rangle $ with
\begin{equation}
N(\vec{n})+N_{\alpha }=N(\vec{n}^{\prime })+N_{\beta }
\end{equation}%
with $N_{\alpha }=1$ for $\alpha =+$ and $N_{\alpha }=0$ for $\alpha =-$.
Obviously, this treatment is a direct generalization of the one in the
development of the GRWA for the TLS-coupled single-mode bosonic field (shown
in appendix A).

Furthermore, as proved in appendix C, we have%
\begin{equation}
H_{{\vec{n}^{\prime }},+}^{{\vec{n}},-}=0\text{ \ \ \ }\mathrm{for}\text{ \ }%
N({\vec{n}})=N({\vec{n}^{\prime }}).  \label{ace}
\end{equation}%
Then the intra-band transition only occurs between the states $|\Psi
_{\alpha ,{\vec{n}}}\rangle $ and $|\Psi _{\alpha ,{\vec{n}}^{\prime
}}\rangle $ with $N({\vec{n}})=N({\vec{n}^{\prime }})$. Therefore, all the
intra-band transitions, which are unreasonably neglected in the adiabatic
approximation, are included in
our GRWA approach through the first term of the right hand side of Eq.~(%
\ref{htg}).
Thus the intrinsic problem of the adiabatic approximation is overcome in our
GRWAapproach. Therefore, our approach is applicable in the
far-off resonance cases with $\Omega <<\omega $.

On the other hand, it is easy to prove that, under the near-resonance
condition $\Omega\sim\omega$ and weak-coupling condition $|\lambda|<<\omega$%
, our GRWAapproach returns to the rotating-wave approximation.
Therefore, as shown in Sec.~III, our approach is applicable in a broad
parameter region with $\Omega\lesssim\omega$.

Finally, we point out that, the approximate Hamiltonian $H_{G}$ in the GRWA
approach in our case can also be re-expressed in a simple form
\begin{equation}
H_{G}=UH_{R}^{\mathrm{RWA}}U^{-1}  \label{ggrwa}
\end{equation}%
with $U$ given by Eq.(\ref{utt1}) and $H_{R}^{\mathrm{RWA}}$ defined as
\begin{equation}
H_{R}^{\mathrm{RWA}}\equiv \sum_{n=0}^{\infty }\left[ \hat{P}_{n}\left(
U^{-1}HU\right) \hat{P}_{n}\right] .  \label{htrrwa}
\end{equation}%
Here the operator $\hat{P}_{n}$ is defined as
\begin{equation}
\hat{P}_{n}=\left\{
\begin{array}{l}
|0\rangle \langle 0|\otimes |g\rangle \langle g|,\ \ \ \ \ \ \ \ \ \ \ \ \ \
\ \ \ \ \ \ \ \ \ \ \ \ \ \ \ \ n=0\nonumber \\
\\
\sum_{\{m_{l}\}}\prod_{l}|m_{l}\rangle _{l}\langle m_{l}|\otimes |g\rangle
\langle g|\delta _{\sum_{l}m_{l},n}+\nonumber \\
\\
\sum_{\{m_{l}\}}\prod_{l}|m_{l}\rangle _{l}\langle m_{l}|\otimes |e\rangle
\langle e|\delta _{\sum_{l}m_{l},n-1},\nonumber \\
\ \ \ \ \ \ \ \ \ \ \ \ \ \ \ \ \ \ \ \ \ \ \ \ \ \ \ \ \ \ \ \ \ \ \ \ \ \
\ \ \ \ \ \ \ \ \ \ \ \ n\neq 0%
\end{array}%
\right.
\end{equation}%
where $|m_{l}\rangle _{l}$ is the Fock state of the $l$-th resonator with $%
m_{l}$ photons. It is apparent that $\hat{P}_{n}$ is the projection operator
to the eigen-space of the total excitation operator
\begin{equation}
|e\rangle \langle e|+\sum_{j=-\infty }^{+\infty }a_{j}^{\dagger }a_{j}
\label{te}
\end{equation}%
with respect to eigen-value $n$. In this sense, similarly to the GRWA for
the TLS-coupled single-mode bosonic field discussed in appendix A, our GRWA
approach can also be considered as \textquotedblleft the rotating-wave
approximation for the rotated Hamiltonian $U^{-1}HU$".

\section{The single-photon scattering amplitudes}

\label{sec3}

In the above section we generalize the GRWA to the hybrid system with a $1$D
resonator array coupled to a single TLS. The single-photon scattering
amplitudes in such a system have been calculated analytically under the
rotating-wave approximation~\cite{zhou08a}. In this section we calculate the
single-photon scattering amplitudes with our GRWA approach which is
applicable in a broader parameter region.

\subsection{The single-photon scattering amplitudes}

The single-photon scattering amplitudes can be extracted from the asymptotic
behavior of the eigen-state of the Hamiltonian $H$, which is approximated as
$H_{G}$ in our GRWA approach. To this end, we need to solve the
eigen-equation
\begin{equation}
H_{G}|\Psi \left( k\right) \rangle =E\left( k\right) |\Psi \left( k\right)
\rangle  \label{ee}
\end{equation}%
with boundary conditions
\begin{eqnarray}
|\Psi \left( k\right) \rangle &=&\left( e^{ik\left( -j\right)
}+r_{k}e^{-ik\left( -j\right) }\right) |1\rangle _{-j}|0\rangle _{j}|\Phi
^{\prime }\left( k\right) \rangle  \notag  \label{bc} \\
&&+t_{k}e^{ikj}|0\rangle _{-j}|1\rangle _{j}|\Phi ^{\prime }\left( k\right)
\rangle +|0\rangle _{-j}|0\rangle _{j}|\Phi \left( k\right) \rangle  \notag
\\
&&  \label{bond}
\end{eqnarray}%
in the limit of $j\rightarrow +\infty $. Here $|0\rangle _{-j}|0\rangle _{j}$
is the vacuum state of the resonator modes in the $j$-th and $-j$-th
resonator, $|1\rangle _{-j}|0\rangle _{j}$ and $|0\rangle _{-j}|1\rangle
_{j} $ are defined as $a_{-j}^{\dagger }|0\rangle _{-j}|0\rangle _{j}$ and $%
a_{j}^{\dagger }|0\rangle _{-j}|0\rangle _{j}$ respectively. $|\Phi
(k)\rangle $ and $|\Phi^{\prime } (k)\rangle $ are the quantum states of the
TLS and other resonators except the $\pm j$-th ones. $r_{k}$ and $t_{k}$ are
the single-photon reflection and transmission amplitudes, or the
single-photon scattering amplitudes.

The physical meaning of the boundary condition (\ref{bond}) can be
understood as follows: For the scattering state with respect to a single
photon input from the left of the $1$D resonator array, there are three
possible relevant states for the $-j$-th and $j$-th resonators with
large $|j|$, i.e., $|0\rangle _{-j}|0\rangle _{j}$, $|1\rangle
_{-j}|0\rangle _{j}$ and $|0\rangle _{-j}|1\rangle _{j}$. Furthermore, the
possibility amplitude with respect to $|1\rangle _{-j}|0\rangle _{j}$ is $%
\left( e^{ik\left( -j\right) }+r_{k}e^{-ik\left( -j\right) }\right) $, since
the photon in the $-j$-th resonator can be either the input one or the
reflected one. Similarly, the possibility amplitude with respect to $%
|0\rangle _{-j}|1\rangle _{j}$ is $t_{k}e^{ikj}$. It is easy to prove that
the boundary condition used in the calculation of the single-photon
scattering state with rotating-wave approximation (Eq.~(5) of Ref.~\cite%
{zhou08a}) can be re-expressed as the one in Eq.~(\ref{bond}).

Usually the expression of $H_{G}$ in Eq.~(\ref{ggrwa}) is complicated and
the eigen-equation (\ref{ee}) is difficult to be solved directly. However,
due to Eq.~(\ref{ggrwa}), the Hamiltonian $H_{G}$ is related to $H_{R}^{%
\mathrm{RWA}}$ through a unitary transformation. Then the eigen-equation (%
\ref{ee}) of $H_{G}$ is equivalent to the one of $H_{R}^{\mathrm{RWA}}$:%
\begin{equation}
H_{R}^{\mathrm{RWA}}|\Psi _{R}\left( k\right) \rangle =E\left( k\right)
|\Psi _{R}\left( k\right) \rangle  \label{nee}
\end{equation}%
and the eigen-state $|\Psi _{R}\left( k\right) \rangle $ of $H_{R}^{\mathrm{%
RWA}}$ is given by
\begin{equation}
|\Psi _{R}\left( k\right) \rangle =U^{-1}|\Psi \left( k\right) \rangle .
\end{equation}%
More importantly, with the help of Eq.(\ref{utt1}) and the fact that $%
\lim_{|j|\rightarrow \infty }\alpha _{j}=0$, the boundary condition (\ref{bc}%
) for $|\Psi \left( k\right) \rangle $ is transformed to the one of $|\Psi
_{R}\left( k\right) \rangle $, i.e., in the limit of $j\rightarrow \infty $
we have%
\begin{eqnarray}
&&|\Psi _{R}\left( k\right) \rangle =\left( e^{ik\left( -j\right)
}+r_{k}e^{-ik\left( -j\right) }\right) |1\rangle _{-j}|0\rangle _{j}|\Phi
_{R}^{\prime }\left( k\right) \rangle  \notag  \label{nbc} \\
&&+t_{k}e^{ikj}|0\rangle _{-j}|1\rangle _{j}|\Phi _{R}^{\prime }\left(
k\right) \rangle +|0\rangle _{-j}|0\rangle _{j}|\Phi _{R}\left( k\right)
\rangle ,
\end{eqnarray}%
where $|\Phi _{R}\left( k\right) \rangle $ and $|\Phi _{R}^{\prime }\left(
k\right) \rangle $ are defined as
\begin{eqnarray}
|\Phi _{R}\left( k\right) \rangle &=&\prod\limits_{i\neq \pm j}\exp [-\alpha
_{i}\sigma _{x}(a_{i}^{\dagger }-a_{i})]|\Phi \left( k\right) \rangle ,
\end{eqnarray}
and
\begin{eqnarray}
|\Phi _{R}^{\prime }\left( k\right) \rangle &=&\prod\limits_{i\neq \pm
j}\exp [-\alpha _{i}\sigma _{x}(a_{i}^{\dagger }-a_{i})]|\Phi ^{\prime
}\left( k\right) \rangle .
\end{eqnarray}%
respectively. Therefore, the scattering amplitudes $r_{k}$ and $t_{k}$ can be obtained
from the solution of the eigen-equation (\ref{nee}) of $H_{R}^{\mathrm{RWA}}$
with boundary conditions (\ref{nbc}).

\subsection{The perturbative approach for the single-phton scattering
amplitudes}

In this subsection we solve the eigen-equation (\ref{nee}) of $H_{R}^{%
\mathrm{RWA}}$ and calculate the single-photon scattering amplitudes. To
this end, we first use the explicit result about the unitary operator $U$
shown in appendix B to calculate the Hamiltonian $U^{-1}HU$. We have the
results as:
\begin{eqnarray}
U^{-1}HU &=&\omega \sum_{j=-\infty }^{\infty }a_{j}^{\dagger }a_{j}-\xi
\sum_{j=-\infty }^{\infty }(a_{j}^{\dagger }a_{j+1}+H.c.)  \notag \\
&&+\frac{\text{$\Omega $}}{2}\left[ \cosh (\sum_{i}2\nu _{i})\sigma
_{z}-i\sinh (\sum_{i}2\nu _{i})\sigma _{y}\right]  \notag \\
&&  \label{htrr}
\end{eqnarray}%
with $\nu _{i}=-\alpha _{i}(a_{i}^{\dagger }-a_{i})$. In principle, we can
derive the explicit expression of $H_{R}^{\mathrm{RWA}}$ with Eqs. (\ref%
{htrrwa}) and (\ref{htrr}). Here, for simplicity, we expand $H_{R}^{\mathrm{%
RWA}}$ as a power series of the parameter $\xi /\omega $ and only keep the
low-order terms under the weak-hopping condition $\xi <<\omega $. Then we
can analytically solve Eq. (\ref{nee}) with the approximated $H_{R}^{%
\mathrm{RWA}}$ and derive the single-photon scattering amplitudes.

Now we calculate the single-photon scattering amplitudes with first-order
approximation where only the $0$-th order and $1$-st order terms of $\xi
/\omega $ are kept in $H_{R}^{\mathrm{RWA}}$. As shown in the following, in
most of the cases, this approximation is enough to give accurate results for
the scattering amplitudes. The straightforward calculation shows that, up to
the first order of $\xi /\omega $, $H_{R}^{\mathrm{RWA}}$ is approximated as
\begin{eqnarray}
H_{R}^{\mathrm{RWA}} &\approx &H_{R}^{\mathrm{RWA}(1)}  \notag \\
&\equiv &\omega \sum_{j}a_{j}^{\dagger }a_{j}-\xi \sum_{j}(a_{j+1}^{\dagger
}a_{j}+h.c.)  \notag \\
&&+\omega _{0g}^{\left( 0\right) }(|0g\rangle \langle 0g|+\sum_{j\neq
0}|1_{j}g\rangle \langle 1_{j}g|)+\omega _{0e}^{\left( 0\right) }|0e\rangle
\left\langle 0e\right\vert  \notag \\
&&+\omega _{1g}^{\left( 0\right) }\left\vert 1_{0}g\right\rangle
\left\langle 1_{0}g\right\vert +J^{\left( 0\right) }\left( \left\vert
0e\right\rangle \left\langle 1_{0}g\right\vert +h.c.\right)  \notag \\
&&+\omega _{1g}^{\left( 1\right) }\left( \left\vert 1_{1}g\right\rangle
\left\langle 1_{0}g\right\vert +\left\vert 1_{-1}g\right\rangle \left\langle
1_{0}g\right\vert +h.c.\right)  \notag \\
&&+J^{\left( 1\right) }\left( \left\vert 0e\right\rangle \left\langle
1_{1}g\right\vert +\left\vert 0e\right\rangle \left\langle
1_{-1}g\right\vert +h.c.\right)  \label{hrwa1}
\end{eqnarray}%
with the parameters
\begin{subequations}
\begin{eqnarray}
\omega _{0g}^{\left( 0\right) } &=&-\omega _{0e}^{\left( 0\right) }=-\frac{%
\Omega }{2}\exp \left( -2\frac{\lambda ^{2}}{\omega ^{2}}\right) ;
\label{omega0g} \\
\omega _{1g}^{\left( 0\right) } &=&-\frac{\Omega }{2}\exp \left( -2\frac{%
\lambda ^{2}}{\omega ^{2}}\right) \left( 1-4\frac{\lambda ^{2}}{\omega ^{2}}%
\right) ; \\
J^{\left( 0\right) } &=&\frac{\Omega \lambda }{\omega }\exp \left( -2\frac{%
\lambda ^{2}}{\omega ^{2}}\right) ; \\
\omega _{1g}^{\left( 1\right) } &=&\frac{2\Omega \lambda ^{2}\xi }{\omega
^{3}}\exp \left( -2\frac{\lambda ^{2}}{\omega ^{2}}\right) ; \\
J^{\left( 1\right) } &=&\frac{\Omega \lambda \xi }{\omega ^{2}}\exp \left( -2%
\frac{\lambda ^{2}}{\omega ^{2}}\right) .  \label{omega1g}
\end{eqnarray}%
In Eq.~(\ref{hrwa1}) the states $|0e\rangle $, $|0g\rangle $, $%
|1_{j}e\rangle $ and $|1_{j}g\rangle $ are defined as $|0\rangle |e\rangle $%
, $|0\rangle |g\rangle $, $|1_{j}\rangle |e\rangle $ and $|1_{j}\rangle
|g\rangle $ respectively, where $|0\rangle $ is the vacuum state of all the
resonators.

\begin{figure*}[tbh]
\includegraphics[bb=13 211 553 628, width=14 cm, clip]{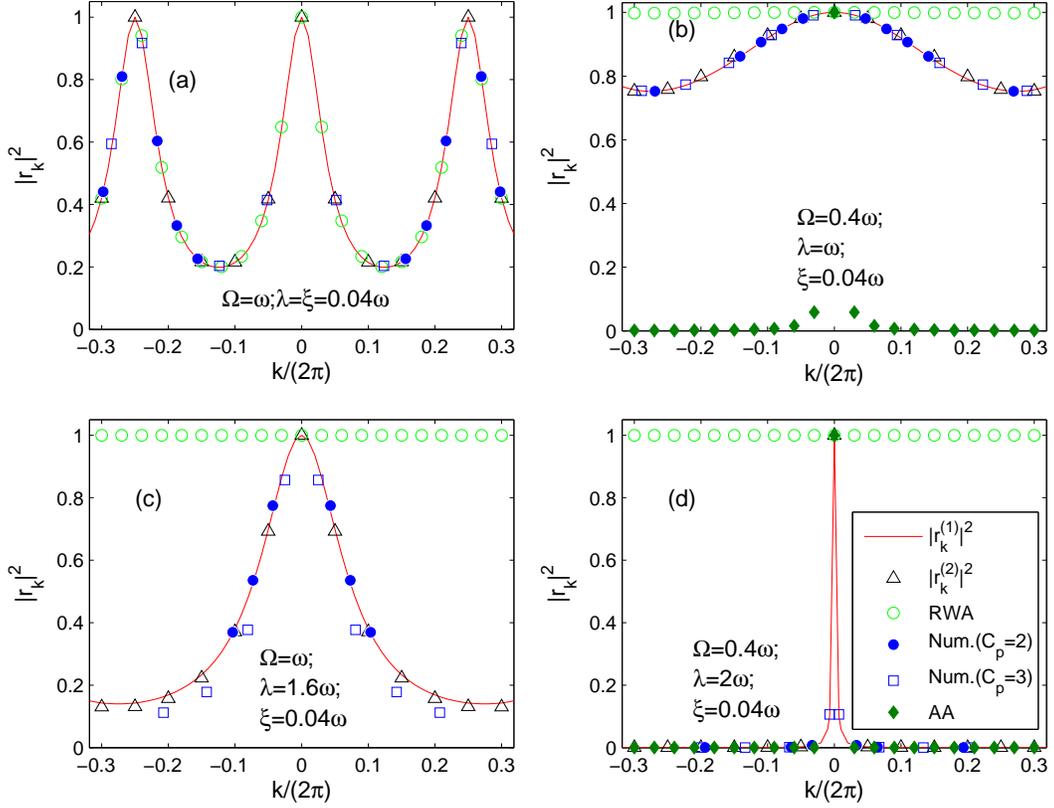}
\caption{(Color online) The single-photon reflection rate $|r_{k}|^{2}$
given by our GRWA approach with the first and second order approximation for
$(\protect\xi /\protect\omega )$, i.e., $|r_{k}^{(1)}|^{2}$ (red solid line)
and $|r_{k}^{(2)}|^{2}$ (black empty triangle), $|r_{k}|^{2}$ from the
rotating-wave approximation (RWA) (green empty circle), $|r_{k}|^{2}$ from
the adiabatic approximation (green filled diamond) and the numerical
calculations with cut-off excitation number $C_{p}=2$ (blue filled circle)
and $C_{p}=3$ (blue empty square). Here we consider the cases of $\protect%
\xi =0.04\protect\omega $ and $\Omega =\protect\omega ,\ \protect\lambda %
=0.04\protect\omega $ (a), $\Omega =0.4\protect\omega ,\ \protect\lambda =%
\protect\omega $ (b), $\Omega =\protect\omega ,\ \protect\lambda =1.6\protect%
\omega $ (c) and $\Omega =0.4\protect\omega ,\ \protect\lambda =2\protect%
\omega $ (d).}
\end{figure*}

The physical meaning of Eq.(\ref{hrwa1}) is very clear. In the $0$-th order
terms of $\xi /\omega $, or the terms proportional to $\omega _{0g}^{\left(
0\right) }$, $\omega _{0e}^{\left( 0\right) }$, $\omega _{1g}^{\left(
0\right) }$ and $J^{\left( 0\right) }$, the effective couplings occur
between the TLS and the photon in the $0$-th resonator in which the TLS is
located. Nevertheless, in the $1$-st order terms proportional to $J^{\left(
1\right) }$, effective couplings appear between the TLS and the modes in the
$\pm 1$-st resonators. These terms imply that the non-rotating-wave effects
from the coupling between the TLS and the $0$-th resonator can indirectly
influence the behavior of the modes in the $\pm 1$-st resonators.
Furthermore, the hopping intensities between the $0$-th and $\pm 1$-st
resonators are also tuned by the terms with $\omega _{1g}^{\left( 1\right) }$%
.

\begin{figure*}[tbh]
\includegraphics[bb=7 207 896 621, width=18 cm, clip]{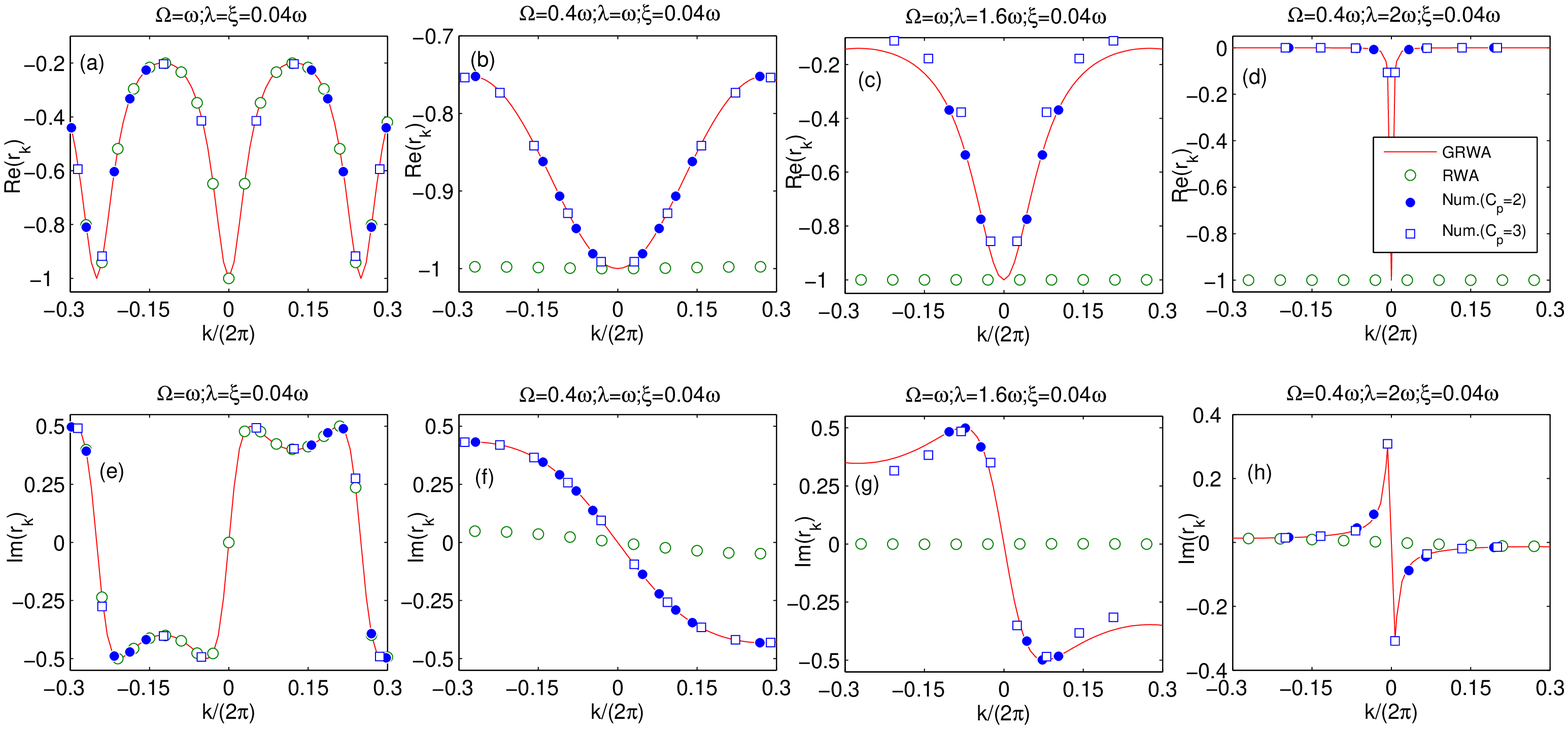}
\caption{(Color online) The real parts (a-d) and imaginary parts (e-h) of
the single-photon scattering amplitude $r_{k}$ given by our GRWA approach
with the first-order approximation for $(\protect\xi /\protect\omega )$,
i.e., $r_{k}^{(1)}$ in Eq.(\protect\ref{rk1}) (red solid line), the
rotating-wave approximation (green empty circle) and the numerical
calculations with cut-off excitation number $C_{p}=2$ (blue filled circle)
and $C_{p}=3$ (blue empty square). Here we consider the cases of $\protect%
\xi =0.04\protect\omega $ and $\Omega =\protect\omega ,\ \protect\lambda %
=0.04\protect\omega $ (a, e), $\Omega =0.4\protect\omega ,\ \protect\lambda =%
\protect\omega $ (b, f), $\Omega =\protect\omega ,\ \protect\lambda =1.6%
\protect\omega $ (c, g) and $\Omega =0.4\protect\omega ,\ \protect\lambda =2%
\protect\omega $ (d, h).}
\end{figure*}

As shown above, the single-photon scattering amplitudes are approximately
derived from the eigen-equation
\end{subequations}
\begin{equation}
H_{R}^{\mathrm{RWA}\left( 1\right) }|\Psi _{R}\left( k\right) \rangle
=E\left( k\right) |\Psi _{R}\left( k\right) \rangle  \label{ee1o}
\end{equation}%
of $H_{R}^{\mathrm{RWA}\left( 1\right) }$ with boundary condition (\ref{nbc}%
). It is apparent that the solution $|\Psi _{R}\left( k\right) \rangle $ of
Eq.(\ref{ee1o}) takes the form%
\begin{equation}
|\Psi _{R}\left( k\right) \rangle =\sum_{j=-\infty }^{+\infty }u_{k}^{\left(
1\right) }(j)\left\vert 1_{j}g\right\rangle +u_{e}^{\left( 1\right)
}\left\vert 0e\right\rangle  \label{ukj1}
\end{equation}%
with the coefficients $u_{k}^{\left( 1\right) }(j)$ given by
\begin{equation}
u_{k}^{\left( 1\right) }(j)=\left\{
\begin{array}{cc}
e^{ikj}+r_{k}^{\left( 1\right) }e^{-ikj}, & j\leqslant -1 \\
u_{k}(0), & j=0 \\
t_{k}^{\left( 1\right) }e^{ikj}. & j\geqslant 1%
\end{array}%
\right. .  \label{nee1}
\end{equation}%
Substituting Eqs.(\ref{ukj1},\ref{nee1}) into Eq.(\ref{ee1o}), we obtain
the linear equations for the reflection amplitude $r_{k}^{\left( 1\right) }$
and transmission amplitude $t_{k}^{\left( 1\right) }$. These equations can
be solved analytically. Then we obtain the scattering amplitudes $%
r_{k}^{\left( 1\right) }$ and $t_{k}^{\left( 1\right) }$ given by the
first-order approximation of $H_{R}^{\mathrm{RWA}}$:
\begin{widetext}
\begin{eqnarray}
r_{k}^{(1)} &=&\frac{e^{ik}\lambda ^{2}\Omega \left\{
\begin{array}{c}
4\lambda ^{2}\xi \Omega ^{2}\cos (k)+2e^{\frac{4\lambda ^{2}}{\omega ^{2}}%
}\omega ^{3}\left[ \omega ^{2}+2\xi \omega \cos (k)-4\xi ^{2}-4\xi ^{2}\cos
\left( 2k\right) \right]  \\
+e^{\frac{2\lambda ^{2}}{\omega ^{2}}}\Omega \left[ 8\lambda ^{2}\xi
^{2}+2\omega ^{2}\xi ^{2}-\omega ^{4}-4\xi \omega \left( 2\lambda
^{2}+\omega ^{2}\right) \cos \left( k\right) +2\xi ^{2}\left( 4\lambda
^{2}+\omega ^{2}\right) \cos \left( 2k\right) \right]
\end{array}%
\right\} }{\left\{
\begin{array}{c}
-4e^{2ik}\lambda ^{4}\xi \Omega ^{3}+e^{\frac{6\lambda ^{2}}{\omega ^{2}}%
}\left( -1+e^{2ik}\right) \xi \omega ^{6}\left[ \omega -2\xi \cos \left(
k\right) \right]  \\
+e^{\frac{4\lambda ^{2}}{\omega ^{2}}}\omega ^{3}\Omega \left[
8e^{3ik}\lambda ^{2}\xi ^{2}+2e^{ik}\lambda ^{2}\left( 4\xi ^{2}-\omega
^{2}\right) +\xi \omega \left( 2\lambda ^{2}+\omega ^{2}\right) -e^{2ik}\xi
\omega \left( 6\lambda ^{2}+\omega ^{2}\right) \right]  \\
+e^{2ik+\frac{\lambda ^{2}}{\omega ^{2}}}\lambda ^{2}\Omega ^{2}\left[ 4\xi
\omega \left( 2\lambda ^{2}+\omega ^{2}\right) +\left( \omega ^{4}-4\xi
^{2}\omega ^{2}-16\lambda ^{2}\xi ^{2}\right) \cos \left( k\right) -i\omega
^{4}\sin \left( k\right) \right]
\end{array}%
\right\} },  \label{rk1} \\
t_{k}^{(1)} &=&r_{k}^{(1)}+1.  \label{tk1}
\end{eqnarray}
\end{widetext}

The above procedure can be straightforwardly generalized to the cases with
high-order approximations of $H_{R}^{\mathrm{RWA}}$. For instance, in the
second-order approximation, $H_{R}^{\mathrm{RWA}}$ is approximated as $%
H_{R}^{\mathrm{RWA}\left( 2\right) }$ which includes the $0$-th order, $1$%
-st order and $2$-nd order terms of $\xi /\omega $. It can be found that in $%
H_{R}^{\mathrm{RWA}\left( 2\right) }$ the TLS is effectively coupled to the $%
0$-th, $\pm 1$-st and $\pm 2$-nd resonators. We can also solve the
eigen-equation of $H_{R}^{\mathrm{RWA}\left( 2\right) }$, and obtain the
analytical expressions of the relevant scattering amplitudes $r_{k}^{\left(
2\right) }$ and $t_{k}^{\left( 2\right) }$. In general, for any integer $n$,
the scattering amplitudes $r_{k}^{\left( n\right) }$ and $t_{k}^{\left(
n\right) }$ from the $n$-th order approximation of $H_{R}^{\mathrm{RWA}}$
can be obtained with the similar approach. In the limit of $n\rightarrow
\infty ,$ the results $r_{k}^{\left( n\right) }$ and $t_{k}^{\left( n\right)
}$ would converge to fixed values $r_{k}$ and $t_{k}$ or the precise values
of the single-photon scattering amplitudes.

\subsection{Results and discussions}

In Figs.~3 and~4, we illustrate the single-photon scattering amplitude $%
r_{k} $ given by our GRWA approach with the first and second order
approximations, i.e., $r_{k}^{(1)}$ and $r_{k}^{(2)}$, the one given by the
rotating-wave approximation and the result from the numerical
diagonalization of the rotated Hamiltonian $U^{-1}HU$. In our numerical
calculations the total excitation defined in (\ref{te}) is cut off at a
given number $C_{p}$ for the Hamiltonian $U^{-1}HU$, and the results with $%
C_{p}=2,\ 3$ are shown in our figures.

In Fig.~3 we calculate the reflection rate $|r_{k}|^{2}$. It is clearly
shown that the results $|r_{k}^{(1)}|^{2}$ and $|r_{k}^{(2)}|^{2}$ from the
first and second order approximation for $(\xi /\omega )$ consist very well
with each other. Therefore, in most of the cases with $|\xi |<<\omega ,$ the
first order approximation is good enough for our GRWA approach.

Furthermore, it is shown that in the case of Fig.~3(a) where the
weak-coupling and near-resonance conditions are satisfied, both the results
from the rotating-wave approximation and our GRWA approach fit well with the
numerical calculations. Nevertheless, in the cases of Figs.~3(b)-3(d) where
the rotating-wave approximation is not applicable, the results from our GRWA
approach also consist significantly well with the numerical calculations.
This observation is further confirmed by Fig.~4 where the real and imaginary
parts of $r_{k}$ given by different approaches are illustrated.

In Figs.~3(b) and 3(d) with $\Omega =0.4\omega $, we also compare our results
with the ones given by the adiabatic approximation. It is shown that as we
argued in Sec. II, the adiabatic approximation may be not applicable even
when $\Omega <<\omega $, while our GRWA approach can still provide
reasonable results.

Therefore, the results in Figs.~3 and 4 show that, our GRWA approach with
the first-order approximation for $(\xi /\omega )$, or our results $%
r_{k}^{(1)}$ and $t_{k}^{(1)}$ in Eqs.~(\ref{rk1}, \ref{tk1}) can be used as
a good analytical approximation for the single-photon scattering amplitudes
in the parameter region with $|\xi |<<\omega $ and $\Omega \lesssim \omega $.

\section{The scattering amplitudes in the strong-coupling case}

\label{sec4}

In the above section we derived the single-photon scattering amplitudes with
our GRWA approach. It is pointed out that, our results in Eqs. (\ref{omega0g}%
-\ref{omega1g}) are applicable for arbitrary large coupling between the TLS
and the photon. Now we consider a special case where the TLS is strongly
coupled to the photon in the resonator array, so that the condition%
\begin{equation}
\frac{\lambda ^{2}}{\omega ^{2}}e^{-2\frac{\lambda ^{2}}{\omega ^{2}}}<<1
\label{cod}
\end{equation}%
is satisfied. We further assume the frequency $\Omega $ of the TLS is equal
to or smaller than the photon frequency $\omega $, i.e., $\Omega \lesssim
\omega $. In this strong-coupling case the expressions in Eqs.~(\ref{omega0g}%
-\ref{omega1g}) can be significantly simplified and then one obtains simple
pictures for both the quantitative calculation and the qualitative
estimation of the single-photon scattering amplitudes.

Under the condition (\ref{cod}), we only keep the leading term proportional
to $\left( \lambda ^{2}/\omega ^{2}\right) \exp \left( -2\lambda ^{2}/\omega
^{2}\right) $ in $\omega _{0g,e}^{(0)}$, $\omega _{1g}^{(0,1)}$ and $%
J^{(0,1)}$ defined in Eqs.~(\ref{omega0g}-\ref{omega1g}). Then we have
\begin{subequations}
\begin{eqnarray}
\omega _{1g}^{(0)} &\approx &\frac{2\Omega \lambda ^{2}}{\omega ^{2}}\exp
\left( -2\frac{\lambda ^{2}}{\omega ^{2}}\right) , \\
\omega _{1g}^{(1)} &\approx &\frac{\xi }{\omega }\omega _{1g}^{(0)}, \\
\omega _{0g,e}^{(0)},\ J^{(0,1)} &\approx &0.
\end{eqnarray}%
Therefore, in Eq.(\ref{hrwa1}) of the Hamiltonian $H_{R}^{\mathrm{RWA}}$, we
only need to keep the first two terms and the terms proportional to $\omega
_{1g}^{(0)}$ and $\omega _{1g}^{(1)}$. This simplification implies that, in
the strong-coupling regime our system is equivalent to the simple 1D
resonator array with the frequency of the 0-th resonator shifted from $%
\omega $ to $\omega +\omega _{1g}^{(0)}$, while the photonic hopping
intensity between the $0$-th resonator and the $\pm $1-st ones shifted from $%
\xi $ to $\xi +\omega _{1g}^{(1)}$.
\begin{figure}[tbp]
\centering
\subfigure{
    \includegraphics[bb=14 404 315 635, width=6.5cm, clip]{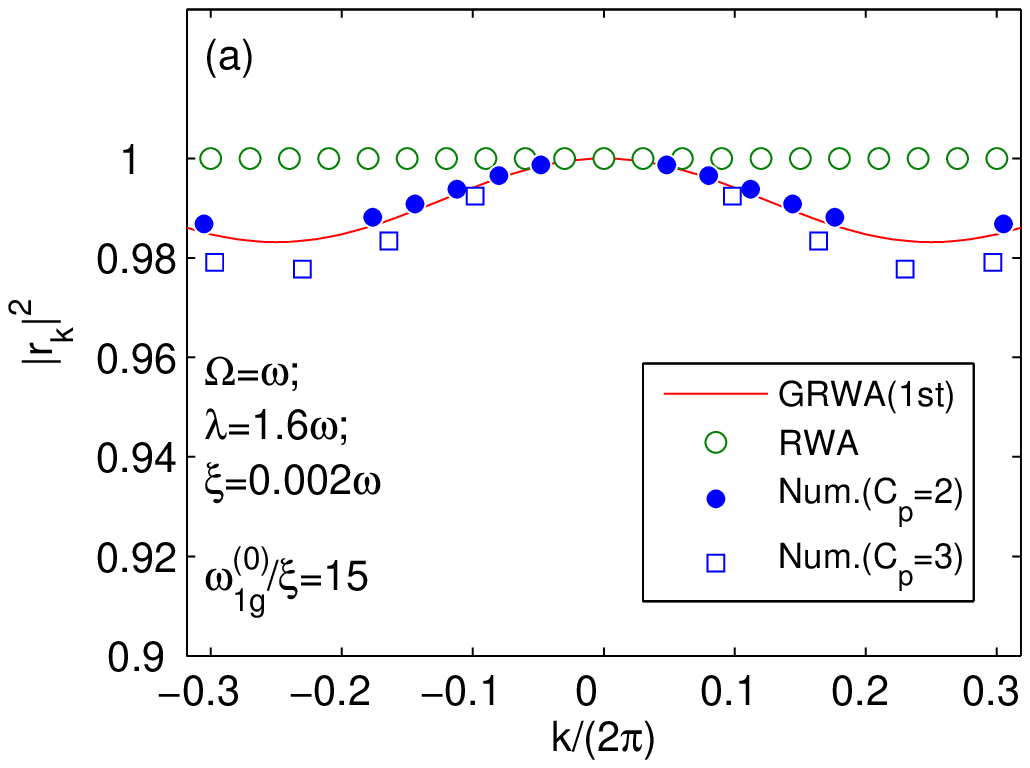}}
\subfigure{
    \includegraphics[bb=1 419 296 654, width=6.5cm, clip]{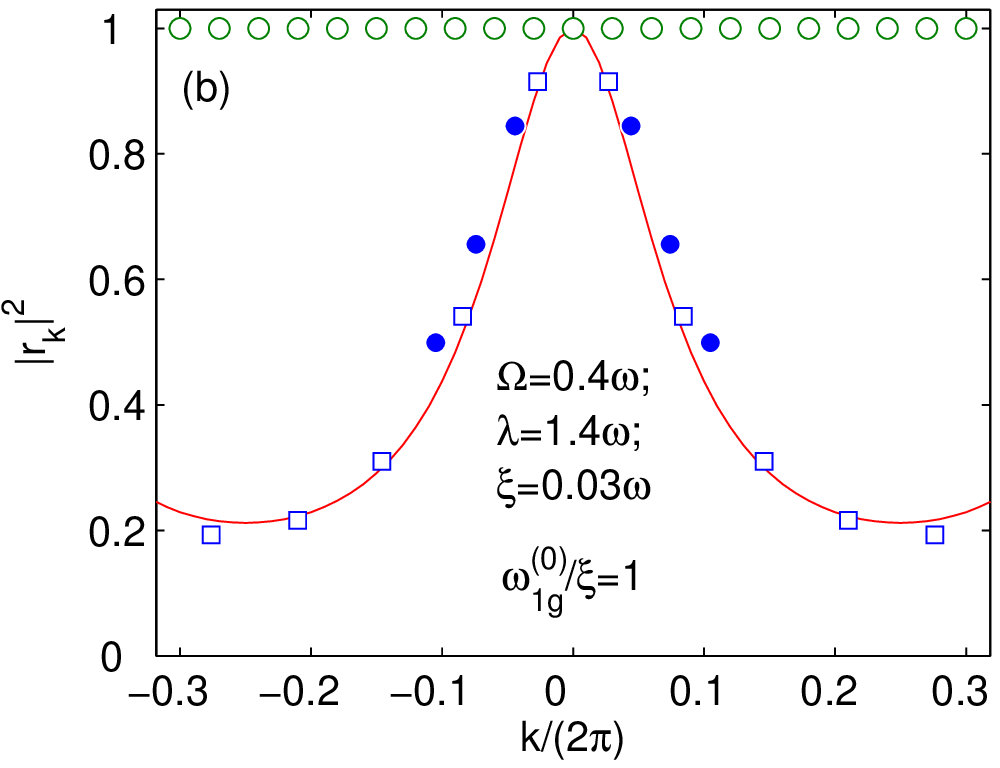}}
\subfigure{
    \includegraphics[bb=1 419 297 654, width=6.5cm, clip]{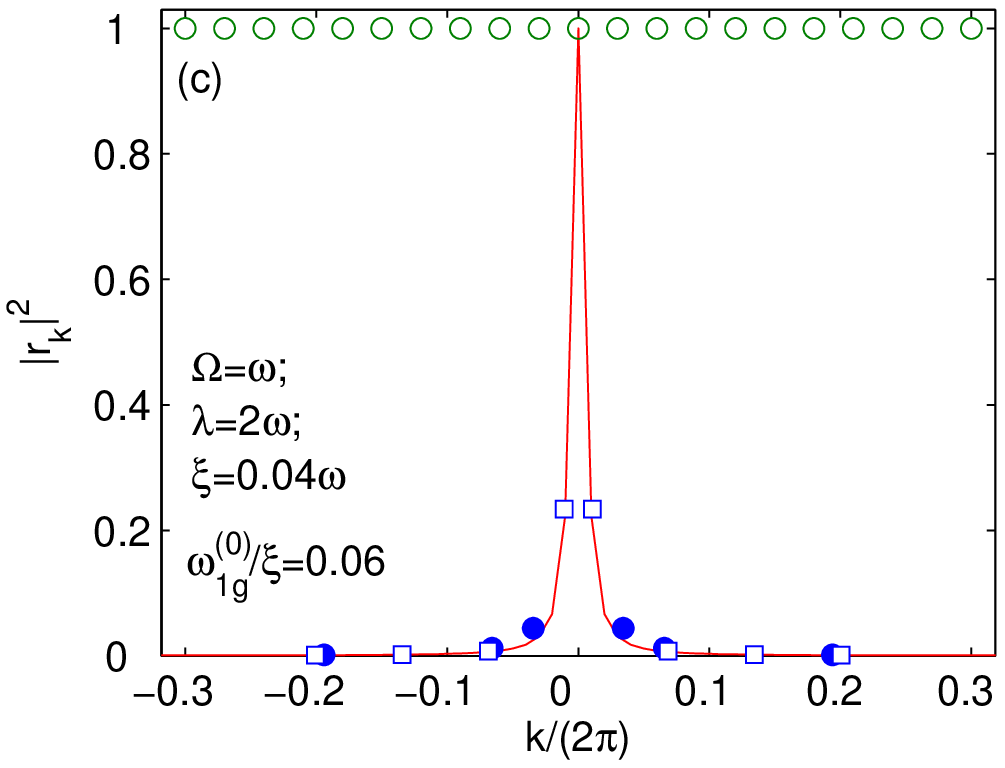}}
\caption{(Color online) The single-photon reflection rate $|r_{k}|^{2}$ in
the cases with strong TLS-photon coupling. Here we show the results given by
Eq.(\protect\ref{rkshift}) (red solid line), rotating-wave approximation
(green empty circle) and the numerical calculations with cut-off excitation
number $C_{p}=2$ (blue filled circle) and $C_{p}=3$ (blue empty square) for
the cases with $\Omega =\protect\omega ,\ \protect\lambda =1.6\protect\omega %
,\ \protect\xi =0.002\protect\omega $ (a), $\Omega =0.4\protect\omega ,\
\protect\lambda =1.4\protect\omega ,\ \protect\xi =0.03\protect\omega $ (b),
$\Omega =\protect\omega ,\ \protect\lambda =2\protect\omega ,\ \protect\xi %
=0.04\protect\omega $ (c). In the three cases we have $\protect\omega %
_{1g}^{(0)}/\protect\xi =15,1,0.06$ respectively.}
\end{figure}
Since we have also assumed $\Omega \lesssim \omega $, it is apparent that $%
\omega _{1g}^{(1)}<<\xi $ in the strong-coupling regime. Then the shift $%
\omega _{1g}^{(1)}$ of the photonic hopping intensity is negligible. We only
need to consider the effect given by the frequency shift $\omega _{1g}^{(0)}$
of the photon in the 0-th resonator. Namely, our system is finally
equivalent to a 1D resonator array, in which the 0-th resonator has the
frequency $\omega +\omega _{1g}^{(0)}$, while all the other resonators have
the same frequency $\omega $. In this case the Hamiltonian $H_{R}^{\mathrm{%
RWA}(1)}$ is approximated as
\end{subequations}
\begin{eqnarray}
H_{R}^{\mathrm{RWA}(1)} &\approx &\omega \sum_{j}a_{j}^{\dagger }a_{j}-\xi
\sum_{j}\left( a_{j+1}^{\dagger }a_{j}+a_{j}^{\dagger }a_{j+1}\right)  \notag
\\
&&+\omega _{1g}^{\left( 0\right) }a_0^{\dagger}a_0,
\end{eqnarray}%
which leads to the single-photon scattering amplitudes
\begin{align}
r_{k}& \approx -\frac{\omega _{1g}^{(0)}}{\omega _{1g}^{(0)}-2i\xi \sin (k)},
\label{rkshift} \\
t_{k}& \approx r_{k}+1.  \label{tkshift}
\end{align}

A straightforward result given by the above expressions of the scattering
amplitudes is that, when the effective frequency shift $\omega _{1g}^{\left(
0\right) }$ of 0-th resonator is much larger than the band width $4\xi $ of
the free Hamiltonian $H_{C}$ of the array of resonators with the same
frequency $\omega $, the 0-th resonator would be far-off detuned with the
photon with any incident momentum $k$, and then every photon would be
reflected. Namely, in such a limit we have $r_{k}\approx 1,\ t_{k}\approx 0.$
Likewise, if the effective frequency shift $\omega _{1g}^{\left( 0\right) }$
is much smaller than $4\xi $, the frequency of the 0-th resonator would be
approximately the same as the other resonators, and then every photon
transmits through the 0-th resonator. In this limit we have $r_{k}\approx 0$
and$\ t_{k}\approx 1$.

In Fig.~5 we plot the photon reflection rate $|r_{k}|^{2}$ in the
strong-coupling cases and compare the results given by our result in Eq.~(%
\ref{rkshift}) and the one from numerical diagonalization of the rotated
Hamiltonian $U^{-1}HU$ with cut-off excitation number $C_{p}=2,\ 3$
respectively. It is clearly shown that our results in Eq. (\ref{rkshift})
fit well with the numerical results. Furthermore, it is illustrated that in
the case of Fig.~5(a) where we have $\omega _{1g}^{\left( 0\right) }/\xi =15$%
, the photon reflection rate $|r_{k}|^{2}$ is almost unit for all the
incident momentum $k$. In the case of Fig.~5(c) with $\omega _{1g}^{\left(
0\right) }/\xi =0.06$, we have $|r_{k}|^{2}\approx 0$ in the region with
non-zero momentum $k$. All these observations are consistent with our above
qualitative analysis.

In the end of this section, we remark that, since all the quantities
defined in Eqs.~(\ref{omega0g}-\ref{omega1g}) exponentially decay to zero
with $(\lambda /\omega )$, for any given values of $\Omega ,\ \omega $ and $%
\xi $, when the TLS-photon coupling intensity $\lambda $ is large enough, we
can always neglect all these parameters and approximate the Hamiltonian $%
H_{R}^{\mathrm{RWA}(1)}$ as the free Hamiltonian $H_{C}$ for the array of
identical resonators. Therefore, when the TLS-photon coupling is strong
enough, the photonic scattering effect becomes negligible and we have $%
r_{k}=0,\ t_{k}=1$ for the photon with any incident momentum $k$.

\section{The GRWA for resonator array with multi TLSs}

\label{sec5}

In the above sections, we generalized the GRWA to the system with $1$D
resonator array coupled to a single TLS, and calculated the single-photon
scattering amplitudes with the GRWA approach. In the end of this paper, we
extend the GRWA to more general cases with $m$ two-level systems coupled to
the resonator array. For simplicity, here we assume each TLS is individually
located in a resonator. Then the Hamiltonian of the total system is written
as
\begin{equation}
H_{M}=H_{C}+H_{AM}+H_{IM}  \label{htm}
\end{equation}%
with the Hamiltonian $H_{C}$ of the resonator defined in Eq. (\ref{hc}), the
Hamiltonian $H_{AM}$ of all the TLSs given by%
\begin{equation}
H_{A}=\frac{\Omega }{2}\sum_{\beta =1}^{m}\sigma _{z}^{\left( \beta \right) }
\end{equation}%
and the interaction Hamiltonian $H_{IM}$ defined as
\begin{equation}
H_{IM}=\lambda \sum_{\beta =1}^{m}\sigma _{x}^{\left( \beta \right) }\left(
a_{c\left( \beta \right) }^{\dagger }+a_{c\left( \beta \right) }\right) .
\end{equation}%
Without loss of generality, here we assume the $\beta $-th TLS is located in
the $c\left( \beta \right) $-th resonator.

In such a general system, we straightforwardly develop the GRWA approach
with the unitary transformation procedure in Sec. II and Sec. III. To this
end, we first write the Hamiltonian $H_{M}$ as
\begin{equation*}
H_{M}=H_{M1}+H_{M2}
\end{equation*}%
with $H_{M1}$ and $H_{M2}$ defined as
\begin{eqnarray}
H_{M1} &=&H_{C}+H_{IM}, \\
H_{M2} &=&H_{AM}.
\end{eqnarray}%
Then we should find a unitary operator $U_{M}$ which can eliminate the linear terms
of $(a_{j},a_{j}^{\dagger })$ in $H_{M1}$ and satisfies%
\begin{eqnarray}
&&U_{M}^{-1}H_{M1}U_{M}  \notag \\
&=&\omega \sum_{j=-\infty }^{\infty }a_{j}^{\dagger }a_{j}-\xi
\sum_{j=-\infty }^{\infty }(a_{j}^{\dagger }a_{j+1}+h.c.)-\mathcal{C}_{M}
\label{utm}
\end{eqnarray}%
with $\mathcal{C}_{M}$ a constant c-number. The analytical calculation of $%
U_{M}$ is obtained in appendix D.

With the operator $U_{M}$, we apply the unitary transformation to the total
Hamiltonian $H$, and make the rotating-wave approximation to the transformed
Hamiltonian $U_{M}^{-1}H_{M}U_{M}$. Finally we perform the inverse unitary
transformation. Then the GRWA Hamiltonian for the resonator array with multi
TLSs is obtained as:%
\begin{eqnarray}
&&H_{M}\approx H_{MG}  \notag \\
&\equiv &U_{M}\left\{ \sum_{n=0}^{\infty }\left[ \hat{P}_{Mn}\left(
U_{M}^{-1}H_{M}U_{M}\right) \hat{P}_{Mn}\right] \right\} U_{M}^{-1},  \notag
\\
&&
\end{eqnarray}%
which is a direct generalization of the result in Eq.~(\ref{ggrwa}). Here $%
\hat{P}_{Mn}$ is the projection operator to the eigen-space of the total
excitation operator
\begin{equation}
\sum_{\beta =1}^{m}|e\rangle ^{\left( \beta \right) }\langle
e|+\sum_{j=-\infty }^{+\infty }a_{j}^{\dagger }a_{j}
\end{equation}%
with respect to eigen-value $n$. It is straightforwardly observed that, in
the case of $\Omega <<\omega ,$ such an approach also includes the
intra-band transitions which are missed in the adiabatic approximation. On
the other hand, under the weak-coupling condition and near-resonance
condition, this approach converges to the rotating-wave approximation.

\section{conclusions}

\label{sec6}

In conclusion we generalize the GRWA to the hybrid system of a 1D
single-mode resonator array coupled to a single TLS, and obtain the
analytical results for the single-photon scattering amplitudes under the
conditions $|\xi |<<\omega $ and $\Omega \lesssim \omega $. It is shown that
in comparison with the rotating-wave approximation, our GRWA approach can
give good results in a much broader parameter region. Especially, in the
far-off resonance case with $\Omega <<\omega $, the adiabatic approximation
is no longer applicable for our current system, while our approach still
works well. We also generalize our GRWA approach to the 1D resonator array
coupled to multi-TLSs.

In this paper, we assume the resonators in the 1D array are single-mode
ones. However, in the resonators used in the experiments, there usually
exist more than one photon modes. Then in the cases with strong TLS-photon
coupling the multi-mode effect may be important. Likewise, it may be also
necessary to go beyond the two-level approximation and include the higher
excited states of the artifical atoms in the strong-coupling cases. These
effects will be discussed in our coming work for the calculation of the
photon scattering in an multi-mode-resonators array or a multi-mode wave
guide beyond the rotating-wave approximation.

\begin{acknowledgments}
This work was supported by National Natural Science Foundation of China {%
under} 
Grants No. 11074305, 10935010, 11074261, 10975181, 11175247, 11174027,
 and the
Research Funds of Renmin University of China (10XNL016).
\end{acknowledgments}

\appendix
\addcontentsline{toc}{section}{Appendices}\markboth{APPENDICES}{}
\begin{subappendices}

\section{GRWA for the TLS-coupled single-mode bosonic field}

In this appendix, we introduce the GRWA approach proposed by Irish in Ref.~%
\cite{irish} in the view of adiabatic approximation. To this end, we
begin with the simple Rabi Hamiltonian for the TLS-coupled single-mode
bosonic field:
\begin{equation}
H_{\mathrm{Rabi}}=\omega a^{\dagger }a+\frac{\Omega }{2}\sigma _{z}+\lambda
\sigma _{x}(a^{\dagger }+a).\label{htls}
\end{equation}%
Here $a$ and $a^{\dagger }$ are the annihilation and creation operators of
the single-mode bosonic field with frequency $\omega $ respectively. $\Omega
$ is the energy level spacing between the excited state $|e\rangle $ and the
ground state $\left\vert g\right\rangle $ of the TLS, and $\lambda $ is the
coupling intensity between the TLS and the bosonic field. The Pauli
operators $\sigma _{z}$ and $\sigma _{x}$ \ are defined in Sec. II.

Since the Hamiltonian $H_{\mathrm{Rabi}}$ does not have simple invariable
subspaces, the exact diagonalization of $H_{\mathrm{Rabi}}$ is rather
complicated~\cite{diag}. However, Jaynes and Cumming showed that~\cite{JC},
under the near-resonance condition
\begin{equation}
\left\vert \omega -\Omega \right\vert <<\left\vert \omega +\Omega
\right\vert
\end{equation}%
and weak coupling condition
\begin{equation}
\left\vert \lambda \right\vert <<\omega ,\Omega,
\end{equation}%
the term $\left\vert e\right\rangle \left\langle g\right\vert a^{\dagger }+h.c$
would be safely neglected. Then the Hamiltonian $H_{\mathrm{Rabi}}$ is
approximated as the $H_{JC}$ which is defined as
\begin{equation}
H_{JC}=\omega a^{\dagger }a+\lambda (|e\rangle \langle g|a+h.c.).\label{hjc}
\end{equation}%
That is the so called rotating-wave approximation. After this approximation,
the Hamiltonian $H_{\mathrm{Rabi}}$ becomes invariable in the
two-dimensional subspaces spanned by the states $|g,n\rangle $ and $%
|e,n-1\rangle $ for $n=$1,2,..., as well as the
one-dimensional subspace spanned by $|g,0\rangle $ and then can be diagonalized easily.

For the convenience of our discussions on the GRWA, here we introduce the
projection operators $\hat{P}_{\mathrm{Rabi}}^{\left( n\right) }$ defined as
\begin{equation}
\hat{P}_{\mathrm{Rabi}}^{\left( n\right) }=\left\{
\begin{array}{l}
\left\vert 0g\right\rangle \left\langle 0g\right\vert, \ \ \ \ \ \ \ \ \ \ \
\ \ \ \ \ \ \ \ \ \ \ \ \ \ \ \ \ \ \ \ \ n=0 \\
\left\vert n,g\right\rangle \left\langle n,g\right\vert +\left\vert
n-1,e\right\rangle \left\langle n-1,e\right\vert, \ n\geq 1\
\end{array}%
\right. .
\end{equation}%
Then the Jaynes-Cumming Hamiltonian $H_{JC}$ in Eq.~(\ref{hjc}) is
re-written as
\begin{equation}
H_{JC}=\sum_{n=0}^{\infty }\hat{P}_{\mathrm{Rabi}}^{\left( n\right) }H_{\rm Rabi}\hat{P}%
_{\mathrm{Rabi}}^{\left( n\right) }.  \label{rwa}
\end{equation}

Now we introduce the GRWA approach, which is developed as an analytical
approximate method to diagonalize the Hamiltonian $H$ in Eq.~(\ref{htls}) in
a broad parameter region where the rotating-wave approximation could be
either applicable or not. The GRWA is closely related to both the
rotating-wave approximation and the adiabatic
approximation ~\cite{BOA1,BOA2,pengBOA} for the TLS-coupled single-mode
bosonic field ~\cite{sun95,irish,irish05} which is used in the case of
far-off resonance
\begin{equation}
\Omega <<\omega .
\end{equation}%
Therefore, before introducing the GRWA, we first introduce the
adiabatic approximation in the system of a TLS and a single-mode
bosonic field\cite{irish,irish05}. In such an approximation, the bosonic
field is considered to be the fast-varying part of the total system and the
TLS is considered as the slowly-varying part. Then the Hamiltonian $H_{%
\mathrm{Rabi}}$ is rewritten as
\begin{equation}
H_{\mathrm{Rabi}}=H_{\mathrm{Rabi}1}+H_{\mathrm{Rabi}2},
\end{equation}%
where
\begin{equation}
H_{\mathrm{Rabi}1}=\omega a^{\dagger }a+\lambda \sigma _{x}(a^{\dagger }+a)
\end{equation}%
is the self-Hamiltonian of the fast-varying part together with the
interaction between the fast-varying part and the slowly-varying one, and
\begin{equation}
H_{\mathrm{Rabi}2}=\frac{\Omega }{2}\sigma _{z}
\end{equation}%
is the self-Hamiltonian of the slowly-varying part.

The Hamiltonian $H_{\mathrm{Rabi}1}$ is easily diagonalized with the
eigenstates:
\begin{equation}
\left\vert \pm ,n\right\rangle =\left\vert \pm \right\rangle \otimes
\left\vert n_{\pm }\right\rangle ,
\end{equation}%
and the relevant eigen-energies:
\begin{equation}
E_{n\pm }=\omega (n-\lambda ^{2}/\omega ^{2}).
\end{equation}%
Here $\left\vert \pm \right\rangle $ are the eigenstates of $\sigma _{x}$
with eigen-values $\pm 1$ and $\left\vert n_{\pm }\right\rangle $ are
defined as%
\begin{equation}
\left\vert n_{\pm }\right\rangle =\exp [\mp \lambda /\omega (a^{\dagger
}-a)]\left\vert n\right\rangle .
\end{equation}%
In the Rabi Hamiltonian the states $%
\left\vert \alpha ,n\right\rangle $ and $\left\vert \alpha ^{\prime
},n^{\prime }\right\rangle $ are coupled by the term $H_{%
\mathrm{Rabi}2}$.

The spirit of the adiabatic approximation is described as follows~%
\cite{BOA1,BOA2,pengBOA}: under the far-off resonance condition $\Omega
<<\omega ,$ the motion of fast-varying part or the bosonic field
adiabatically follows the slowly-varying part or the TLS, and can be frozen
in the adiabatic branches with fixed quantum number $n$, or the
two-dimensional subspaces spanned by $|+,n\rangle $ and $|-,n\rangle $ for $%
n=1,2,...$. Then we neglect the $H_{\mathrm{Rabi}2}$-induced transitions
between the states $\left\vert \alpha ,n\right\rangle $ and $\left\vert
\alpha ^{\prime },n^{\prime }\right\rangle $ with $n\neq n^{\prime }$. Then
the eigen-states and eigen-energies of $H$ are approximated as:
\begin{equation}
|\Psi _{\pm ,n}\rangle =\frac{1}{\sqrt{2}}%
(|+,n\rangle \pm |-,n\rangle ),
\end{equation}%
and
\begin{equation}
E_{\pm ,n}=\pm \frac{\Omega }{2}\langle
n_{-}|n_{+}\rangle +\omega (n-\lambda ^{2}/\omega ^{2}).
\end{equation}%
respectively.

Now we introduce the GRWA. In the \textquotedblleft adiabatic basis"
$
\{|\Psi _{\pm ,n}\rangle \},
$
the Hamiltonian $H$ is rewritten as
\begin{equation}
H_{\mathrm{Rabi}}=\sum_{n,n^{\prime }}\sum_{\alpha ,\alpha ^{\prime }=\pm
}\left( H_{\mathrm{Rabi}}\right) _{\alpha ,n}^{\alpha ^{\prime },n^{\prime
}}|\Psi _{\alpha ,n}\rangle \langle \Psi
_{\alpha ^{\prime },n^{\prime }}|
\end{equation}%
with the matrix elements
\begin{eqnarray}
\left( H_{\mathrm{Rabi}}\right) _{\alpha
,n}^{\alpha ^{\prime },n^{\prime }}=\langle \Psi _{\alpha ,n}|H_{\mathrm{Rabi}}|\Psi _{\alpha ^{\prime },n^{\prime
}}\rangle .
\end{eqnarray}
In the GRWA, the Rabi Hamiltonian $H_{\mathrm{Rabi}}$ is approximated as
$H_{RG}$ which is defined as
\begin{eqnarray}
H_{RG} &=&\sum_{n=0}^{+\infty }\sum_{\alpha =\pm }\left( H_{\mathrm{Rabi}%
}\right) _{\alpha ,n}^{\alpha ,n}|\Psi _{\alpha ,n}\rangle \langle \Psi _{\alpha ,n}|
\notag  \label{hgrwa} \\
&&+\sum_{n=1}^{\infty }\left( H_{\mathrm{Rabi}}\right) _{-,n}^{+,n-1}|\Psi
_{-,n}\rangle \langle \Psi _{+,n-1}|+h.c..  \notag \\
&&
\end{eqnarray}%
Namely, only the matrix elements $\left( H_{\mathrm{Rabi}%
}\right) _{\alpha ,n}^{\alpha ^{\prime },n^{\prime }}$ inside each
two-dimensional subspaces spanned by the states $|\Psi _{-,n}\rangle $ and $|\Psi _{+,n-1}
\rangle $ as well as the one in the one-dimensional subspace
spanned by the state $|\Psi _{-,0}\rangle $ are kept in the GRWA.
In other words, in the GRWA one takes into account only the quantum transitions between
the states $|\Psi _{\alpha ,n}\rangle $ and $%
|\Psi _{\beta ,n^{\prime }}\rangle $ with
\begin{equation}
n+N_{\alpha }=n^{\prime }+N_{\beta }
\end{equation}%
with $N_{\alpha }=1$ for $\alpha =+$ and $N_{\alpha }=0$ for $\alpha =-$.
Then, similarly as in the rotating-wave approximation, the
Hamiltonian is reduced into a $2\times 2$ blocked diagonal matrix in the GRWA.

It can be shown that, under the weak-coupling and near-resonance conditions,
the GRWA returns to the rotating-wave approximation. On the other hand,
under the far-off resonance condition, the results given by the GRWA
converges to the one from adiabatic approximation. Therefore, the GRWA
smoothly connects the ordinary rotating wave approximation and the
adiabatic approximation, and then works well in a more broad
parameter regime, especially the region with strong TLS-photon coupling and
far-off resonant $\Omega \lesssim \omega $ (see, e.g., Ref.~\cite{irish} and
Fig.~8 of Ref.~\cite{grifoni10}).

In the end of this appendix, we emphasis that, the Hamiltonian $H_{RG}$
defined in (\ref{hgrwa}) could also be re-written as~\cite{irish}
\begin{equation}
H_{RG}=U_{R}\left\{ \sum_{n=0}^{\infty }\left( \hat{P}_{n}\left(
U_{R}^{-1}H_{\mathrm{Rabi}}U_{R}\right) \hat{P}_{n}\right) \right\}
U_{R}^{-1}\label{hrrwa}
\end{equation}%
with the unitary transformation $U_{R}$ defined as
\begin{equation}
U_{R}=\exp \left[ -\frac{\lambda }{\omega }\sigma _{x}(a^{\dagger }-a)\right]
.
\end{equation}%
Comparing Eq.~(\ref{hrrwa}) and Eq.~(\ref{rwa}), one can find that the GRWA is
nothing but the \textquotedblleft rotating-wave approximation for the
rotated Hamiltonian $U_{R}^{-1}H_{\mathrm{Rabi}}U_{R}$".

\section{The unitary transformation $U$}

In this appendix, we calculate the unitary transformation $U$ determined by
Eq.~(\ref{utt1}) and Eq.~(\ref{alphaj1}). It is obvious that
 $U$ can be expressed as the product of the
displacement operators for each resonator mode, i.e., we have%
\begin{equation}
U=\prod\limits_{j=-\infty }^{+\infty }\exp [\alpha _{j}\sigma
_{x}(a_{j}^{\dagger }-a_{j})].\label{uuua}
\end{equation}
Then the Hamiltonian $H_{1}$ is transformed into
\begin{eqnarray}
&&U^{-1}H_{1}U  \notag  \label{res} \\
&=&H_{C}+\sum_{j\neq 0}\left[ \omega \alpha _{j}-\xi (\alpha _{j+1}+\alpha
_{j-1})\right] \sigma _{x}(a_{j}^{\dagger }+a_{j})  \notag \\
&&+\left[ \omega \alpha _{0}-\xi (\alpha _{1}+\alpha _{-1})+\lambda \right]
\sigma _{x}(a_{0}^{\dagger }+a_{0})-\mathcal{C}.  \notag \\
&&
\end{eqnarray}%
The exact expression of $\mathcal{C}$ is not required here. Comparing the
result Eq.~(\ref{res}) with Eq.~(\ref{ut2}), we get the equations for the
parameters $\left\{ \alpha _{j}\right\} $:
\begin{eqnarray}
\omega \alpha _{j}-\xi (\alpha _{j+1}+\alpha _{j-1}) &=&0,j=\pm 1,\pm 2,...
\label{e1} \\
\omega \alpha _{0}-\xi (\alpha _{1}+\alpha _{-1})+\lambda  &=&0,j=0.
\label{e2}
\end{eqnarray}

Now we solve Eqs.~(\ref{e1}, \ref{e2}) with two steps. First, we introduce a
cut-off for the equations at $j=\pm N$, and solve the equations
\begin{eqnarray}
&&\omega \alpha _{j}^{\left( N\right) }-\xi \left( \alpha _{j+1}^{\left(
N\right) }+\alpha _{j-1}^{\left( N\right) }\right) =0,\ j=\pm 1,...,\pm (N-1)
\nonumber \\
\label{ea1}\\
&&\omega \alpha _{0}^{\left( N\right) }-\xi \left( \alpha _{1}^{\left(
N\right) }+\alpha _{-1}^{\left( N\right) }\right) +\lambda =0,\ j=0,
\label{ea2} \\
&&\alpha _{N}^{\left( N\right) }=\alpha _{-N}^{\left( N\right) } =0.
\label{ea3}
\end{eqnarray}%
A straightforward calculation shows that
\begin{eqnarray}
\alpha _{j}^{\left( N\right) } &=&\left[ \left( \frac{\omega }{2\xi }-\frac{%
\xi }{\omega _{1}}\right) \alpha _{0}^{\left( N\right) }+\frac{\lambda }{%
2\xi }\right] \frac{(\frac{\xi }{\omega _{1}})^{|j|-1}-(\frac{\omega _{1}}{%
\xi })^{|j|+1}}{1-(\frac{\omega _{1}}{\xi })^{2}}  \notag \\
&&+(\frac{\xi }{\omega _{1}})^{\left\vert j\right\vert }\alpha _{0}^{\left(
N\right) }  \label{diedai}
\end{eqnarray}%
with $\omega _{1}$ given by
\begin{eqnarray}
\omega _{1}=\frac{1}{2}\left( \omega +\sqrt{\omega ^{2}-4\xi ^{2}}\right) .\label{omega1}
\end{eqnarray}%
Therefore, under the weak-hopping assumption $\left\vert \xi \right\vert
<<\omega $, we have $\xi/\omega_1<1$. Substituting Eq.~(\ref{ea3}) into Eq.~(\ref%
{diedai}), we obtain%
\begin{eqnarray}
\alpha _{0}^{\left( N\right) }=\frac{-\lambda \omega _{1}\left[ (\frac{%
\omega _{1}}{\xi })^{2|N|}-1\right] }{-\omega \omega _{1}+2\omega
_{1}^{2}-2\xi ^{2}(\frac{\omega _{1}}{\xi })^{2|N|}+\omega \omega _{1}(\frac{%
\omega _{1}}{\xi })^{2|N|}}\nonumber\\
\end{eqnarray}%
and then we obtain all $\alpha _{j}^{\left( N\right) }$ from Eq.~(\ref%
{diedai}).

Second, we consider the solutions of Eqs.~(\ref{ea1}-\ref{ea3}) in limit of $%
N\rightarrow \infty $ as a trial solution of Eqs.~(\ref{e1}, \ref{e2}):
\begin{eqnarray}
\alpha _{j}=\lim_{N\rightarrow \infty }\alpha _{j}^{\left( N\right) }=\frac{%
\lambda \omega _{1}}{2\xi ^{2}-\omega \omega _{1}}(\frac{\xi }{\omega _{1}}%
)^{|j|}.  \label{alphaj}
\end{eqnarray}%
Substituting Eq.~(\ref{alphaj}) into Eqs.~(\ref{e1}, \ref{e2}), we find that the
latter ones are exactly satisfied. Therefore, $\left\{ \alpha _{j}\right\} $
from Eq.~(\ref{alphaj}) are the realistic solutions of Eqs.~(\ref{e1}, \ref{e2}%
). Then the unitary transformation $U$ defined in Eq.~(\ref{ut2}) takes the
form Eq.~(\ref{utt1}) with $\alpha _{j}$ given by Eq.~(\ref{alphaj1}).

We emphasis that, as shown in Eq.~(\ref{uuua}), $U$ is the product of the
displacement operators for each resonator, the magnitude of the
displacement for the mode in the $j$-th resonator is described by $%
\left\vert \alpha _{j}\right\vert $. Furthermore, since $\xi /\omega _{1}<1$%
, the result Eq.~(\ref{alphaj}) implies that $\left\vert \alpha _{j}\right\vert $
exponentially decays with $\left\vert j\right\vert $. Then we have
\begin{equation}
\lim_{\left\vert j\right\vert \rightarrow \infty }\exp [\alpha _{j}\sigma
_{x}(a_{j}^{\dagger }-a_{j})]=1  \label{lim}
\end{equation}%
Therefore, for the resonators which are far away from the TLS, the relevant
displacements in $U$ would be negligible, this is consistent with our
above considerations.

\section{The proof of Eq.~(\ref{ace})}

In this appendix, we prove Eq.~(\ref{ace}) in our maintext. To this end, we
first define the operator $\mathcal{U}_{k}\left[ \beta_k \right]$ and the state $|n\left(
k\right) \rangle $ for the $1$D resonator array as
a function of the number $\beta_k$:%
\begin{equation}
\mathcal{U}_{k}\left[ \beta_k \right] =\exp \left\{ -2\beta_k \left[ A\left( k\right) ^{\dagger }-A\left( k\right) \right]
\right\}
\end{equation}%
and
\begin{equation}
|n\left( k\right) \rangle =\frac{1}{\sqrt{n\left( k\right) }}A\left(
k\right) ^{\dagger n\left( k\right) }|0\rangle
\end{equation}%
respectively. Here $|0\rangle $ is the vacuum state of the resoator array, $%
A\left( k\right) ^{\dagger }$ is defined in Eq.~(\ref{bigak}).
 We further define the
function $f(\vec{\beta})$ as
\begin{equation}
f(\vec{\beta})=\prod_{k}\langle n\left( k\right) |\mathcal{U}_{k}\left[
\beta_k\right] |n^{\prime }\left( k\right) \rangle
\end{equation}%
with%
\begin{equation}
\vec{\beta}=\left(\beta_{k_1} ,\beta_{k_2} ,...\right).
\end{equation}%

The straightforward calculation shows that %
\begin{widetext}
\begin{eqnarray}
f(\vec{\beta}) &=&\prod_{k}\langle n\left( k\right) |\mathcal{U}_{k}\left[
\beta_k\right] |n^{\prime }\left( k\right) \rangle   \notag \\
&=&\prod_{k}e^{-\left\vert 2\beta_k\right\vert ^{2}/2}\langle
n\left( k\right) |e^{-2\beta_kA\left( k\right) ^{\dagger
}}e^{2\beta_kA\left( k\right) }|n^{\prime }\left( k\right)
\rangle   \notag \\
&=&\prod_{k}\left( e^{-\left\vert 2\beta_k\right\vert
^{2}/2}\sum_{m=\mathrm{Max}\left[ 0,n^{\prime }\left( k\right) -n\left(
k\right) \right] }^{n^{\prime }\left( k\right) }\frac{\left[ 2\beta \left(
k\right) \right] ^{m}\left[ -2\beta_k\right] ^{n\left(
k\right) -n^{\prime }\left( k\right) +m}}{m!\left[ n\left( k\right)
-n^{\prime }\left( k\right) +m\right] !}\langle n\left( k\right) |A\left(
k\right) ^{\dagger m}A\left( k\right) ^{m}|n^{\prime }\left( k\right)
\rangle \right),\notag \\
\end{eqnarray}%
\end{widetext}
which gives%
\begin{eqnarray}
f(-\vec{\beta}) &=&\left( -1\right) ^{N\left( \vec{n}\right) -N\left( \vec{n}%
^{\prime }\right) }f(\vec{\beta})
\end{eqnarray}
or
\begin{eqnarray}
f(-\vec{\beta})&=&f(\vec{\beta})\ \ \ \ \ \ \ \ \ \ \  {\rm for}\ \  N\left( \vec{n}\right)=N\left( \vec{n}^{\prime }\right).\label{aaa}
\end{eqnarray}%

On the other hand, with the above definitions and straightforward calculations, we have%
\begin{eqnarray}
\langle +,\vec{n}|-,\vec{n}^{\prime }\rangle  &=&f(\vec{\beta}_0); \\
\langle -,\vec{n}|+,\vec{n}^{\prime }\rangle  &=&f(-\vec{\beta}_0),
\end{eqnarray}%
where $|\pm ,\vec{n}\rangle $ are the eigen-states of the Hamiltonian $H_{1}$
defined in Sec. IIB and $\vec{\beta}_0=(\beta_{0k_1},\beta_{0k_2},....)$.
Here we have
\begin{equation}
\beta_{0k}=\frac{1}{\sqrt{2\pi }}\sum_{l=-\infty }^{+\infty
}\alpha _{l}e^{ikl}
\end{equation}%
with $\alpha _{l}$ defined in Eq.~(\ref{alphaj1}). Then using Eq.~(\ref{a6}) and Eq.~(\ref{hme}), we
re-write the matrix element $H_{\vec{n}^{\prime },-}^{\vec{n},+}$ as
\begin{eqnarray}
H_{\vec{n}^{\prime },-}^{\vec{n},+} &=&\langle \Psi _{+,\vec{n}}|H|\Psi _{+,\vec{n}}\rangle   \notag \\
&=&\langle \Psi _{+,\vec{n}}|H_{2}|\Psi _{-,%
\vec{n}^{\prime }}\rangle   \notag \\
&=&\frac{\Omega }{4}\left( \langle +,\vec{n}|-,\vec{n}^{\prime }\rangle
-\langle -,\vec{n}|+,\vec{n}^{\prime }\rangle \right)   \notag \\
&=&\frac{\Omega }{4}\left[ f(\vec{\beta}_0)-f(-\vec{\beta}_0)\right]
\end{eqnarray}%
Therefore, our above result in Eq.~(\ref{aaa}) directly leads to Eq.~(\ref{ace}).

\section{The unitary operator $U_{M}$}

In this appendix, we calculate the unitary operator $U_{M}$ defined in (\ref{utm}).
Similar as in appendix B, it is easy to prove that $U_{TM}$ is the product
of the displacement operators of each resonator mode:%
\begin{equation}
U_{M}=\prod_{j=-\infty }^{+\infty }\exp [\alpha _{j}^{M}\sigma
_{x}(a_{j}^{\dagger }-a_{j})].  \label{utm2}
\end{equation}%
To derive the expression of $\alpha _{j}^{M}$, we define the column vector
\begin{equation}
\vec{\alpha}=\left( ...\alpha _{-1}^{M},\alpha _{0}^{M},\alpha
_{1}^{M},...\right) ^{T}.
\end{equation}%
Then the straightforward calculation shows that, the condition (\ref{utm})
is equivalent to the equation%
\begin{equation}
\omega \vec{\alpha}-\xi K\vec{\alpha}=\vec{\Lambda}.  \label{eqa}
\end{equation}%
Here $K$ is a square matrix with the element $K_{ij}$ in the $i$-th row and $%
j$-th column given by
\begin{equation}
K_{ij}=\delta _{i,j+1}+\delta _{i,j-1}
\end{equation}%
In Eq.~(\ref{eqa}), $\vec{\Lambda}$ is a constant column vector with the $j$%
-th component $\Lambda _{j}$ defined as%
\begin{equation}
\Lambda _{j}=\lambda \sum_{\beta =1}^{m}\delta _{j,c\left( \beta \right) }.
\end{equation}%
Therefore, we formally have the expression of $\vec{\alpha}$:%
\begin{equation}
\vec{\alpha}=\frac{1}{\omega -\xi K}\vec{\Lambda}.  \label{af}
\end{equation}

Furthermore, we notice that the matrix $\omega -\xi K$ is diagonalized as%
\begin{equation}
\omega -\xi K=\int dk\left( \omega -2\xi \cos k\right) \vec{v}(k)\vec{v}(k)%
^{\dagger }  \label{dk}
\end{equation}%
with the $j$-th component of the column vector $\vec{v}(k)$ satisfies%
\begin{equation}
v_{j}(k)=\frac{1}{\sqrt{2\pi }}e^{ikj}.
\end{equation}%
Then we have
\begin{equation}
\frac{1}{\omega -\xi K}=\int dk\frac{\vec{v}(k)\vec{v}(k)^{\dagger }}{\omega -2\xi
\cos k}.  \label{ki}
\end{equation}%
Substituting (\ref{ki}) into (\ref{af}), we get the expression of $\alpha
_{j}^{M}$:%
\begin{equation}
\alpha _{j}^{M}=\frac{\lambda }{2\pi }\sum_{\beta =1}^{m}\int dk\frac{\exp %
\left[ ik\left( j-c\left( \beta \right) \right) \right] }{\omega -2\xi \cos k%
}.  \label{ajm}
\end{equation}%
In the case of a single TLS located the $0$-th resonator, we have $m=1$ and $%
c\left( 1\right) =0$. Then $\alpha _{j}^{M}$ is expressed as%
\begin{equation}
\alpha _{j}^{M}=\frac{\lambda }{2\pi }\int dk\frac{\exp \left[ ikj\right] }{%
\omega -2\xi \cos k}.
\end{equation}%
On the other hand, in such a single-TLS case, the value of $\alpha _{j}^{M}$
is also given by (\ref{alphaj}). Therefore, we have
\begin{equation}
\frac{\lambda }{2\pi }\int dk\frac{\exp \left[ ikj\right] }{\omega -2\xi
\cos k}=\frac{\lambda \omega _{1}}{2\xi ^{2}-\omega \omega _{1}}(\frac{\xi }{%
\omega _{1}})^{|j|}  \label{ss}
\end{equation}%
with $\omega _{1}$ defined in Sec.II.B. Substituting (\ref{ss}) into (%
\ref{ajm}), we finally obtain%
\begin{equation}
\alpha _{j}^{M}=\sum_{\beta =1}^{m}\frac{\lambda \omega _{1}}{2\xi
^{2}-\omega \omega _{1}}(\frac{\xi }{\omega _{1}})^{|j-c\left( \beta \right)
|}.
\end{equation}%
Therefore, we get the analytical expression of the unitary operator $U_{M}$
defined in Eqs.~(\ref{utm}, \ref{utm2}).

\end{subappendices}

\end{document}